First Monday, Volume 17, Number 6 - 4 June 2012

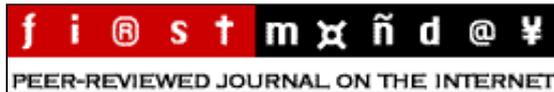

# The relationship between Internet user type and user performance when carrying out simple vs. complex search tasks
## by Georg Singer, Pille Pruulmann-Vengerfeldt, Ulrich Norbisrath, and Dirk Lewandowski

## Abstract
It is widely known that people become better at an activity if they perform this activity long and often. Yet, the question is whether being active in related areas like communicating online, writing blog articles or commenting on community forums have an impact on a person's ability to perform Web searches, is still unanswered. Web searching has become a key task conducted online; in this paper we present our findings on whether the user type, which categorises a person's online activities, has an impact on her or his search capabilities. We show (1) the characteristics of different user types when carrying out simple search tasks; (2) their characteristics when carrying out complex search tasks; and, (3) the significantly different user type characteristics between simple and complex search tasks. The results are based on an experiment with 56 ordinary Web users in a laboratory environment. The Search–Logger study framework was used to analyze and measure user behavior when carrying out a set of 12 predefined search tasks. Our findings include the fact that depending on task type (simple or complex) significant differences can be observed between users of different types.

**Contents**

**Introduction**

More and more people are going online. They use the Internet for all kinds of tasks. They visit social networking sites like Facebook, communicate with friends and family, take part in forum discussions, do their banking online, read the news, book trips, write blog articles, look for information to make decisions, and stay or become informed (Head and Eisenberg, 2011). As research shows (*e.g.*, Brandtzæg, *et al.*, 2011), people with the same level of Internet access can use the Internet in fundamentally different ways. Together with the enormous success of the Internet and its unprecedented growth into all areas of our daily routines, the amount of information available through this medium is still growing rapidly (Gantz, *et al.*, 2008). The Internet, which used to be considered a fairly 'monolithic' medium, has long become diversified, and understanding Internet user types is an important aspect in that diversification (Pruulmann–Vengerfeldt, 2006; Horrigan, 2007; Pruulmann–Vengerfeldt and Reinsalu, 2009; Meyen, *et al.*, 2010).

Having access to the right information is a key requirement for many of the tasks that people carry out online. Web search engines have become the tools of choice to find information on the Internet (Hargittai, 2004). People use search engines to look up simple information, for example when their favourite pop star was born, to find a recipe for a certain dish or to look up a certain location on an online map. Head and Eisenberg (2011) state that, based on a study of 8,353 college students, 95 percent used search engines to find that kind of information.

However, such simple searches are only one kind of task, which are complemented by a growing number of more complex tasks that involve decision–making, learning and investigating. Many of these more complex tasks are not as well supported by Web search engines as their simple counterparts (White and Roth, 2009). One side of the coin is related to search engines; however, the other side relates to the users and their search capabilities. Among the three information types (problem information, domain information, problem–solving information) that are usually needed to carry out search tasks, problem–solving information relates to experience–based aspects like prior knowledge of how a similar problem was solved (Byström and Järvelin, 1995).

Prior Web experience and knowledge might have an impact on a person's efficacy in situations where they are faced with, for example, a decision task in which people compiling a whole lot of information from various sources discover unknown aspects of the decision task that need separate reviewing and synthesizing into a single document (White, *et al.*, 2008; Singer, *et al.*, 2012a). In general, not very much is known about users' behavior when using Web search engines to carry out tasks. This is especially true in case of complex search tasks that can go over long time spans and are hard to measure. And even less is known about how prior Web experience, as well as the type and quality of people's online activities, influence their behavior when carrying out aforementioned complex tasks online. In this article, we mainly rely on our previous work when distinguishing Internet user types. In general, Kalmus, *et al.* (2009) divide Internet users into *information seekers*, *communication seekers*, and and *entertainment seekers*. Adding the active–passive scale to the discussion, we find that there are six basic Internet user types that we have used in our study as indicators of Internet use (Pruulmann–Vengerfeldt, 2006; Pruulmann–Vengerfeldt and Reinsalu, 2009).

In this paper, we present the findings of a study examining how the Internet user type is related to the ability to carry out complex search tasks with Web search engines. We also show the differences between certain user types and discuss the noticeable differences. To be able to analyze online behavior, all users' activity was logged using the Search–Logger tool (Singer, *et al.*, 2011). The study was conducted in a laboratory environment. The user sample consisted of 56 ordinary Web users (with diverse backgrounds) whom we asked to carry out 12 predefined search tasks with Web search tools of their choice (search engines and other online information tools like Wikipedia).

This paper provides a unique combination of Internet user typologies and search engine use practices assuming that most of today's Internet users are fairly active in searching different kinds of information online (Horrigan, 2007). However, little is known about whether their search success and search practices are related to their prior experience and their predominant Internet use, *i.e.*, whether they, for example, mainly run a blog, search for information or communicate online. The results of this "study focus mainly on understanding which group is better able to obtain correct results to simple or complex search problems.

The rest of this paper is structured as follows: Firstly, we review the literature on Internet user types, and on simple, complex, and exploratory searches. Secondly, we state some research questions. Thirdly, we describe our methods, followed by the results. The results are discussed, and in the conclusions section we sum up the outcomes and limitations of our research and give some directions for future research.

---

### Literature review

*Internet user types*

As the Internet has become widely adopted by a wide range of different users (younger and older people, tech–savvy people and beginners, professionals and fun users), there is an increasing need to classify users. In doing so, it becomes possible to understand different uses, maximize the possible support efforts and target content to particular needs. These user typologies can be performed using different approaches — according to their prior experience (*e.g.*, how long has a person been an Internet user) combined with intensity (*e.g.*, how often and how long on any given day do they use the Internet), as per Howard, *et al.* (2001) in their classic study of American Internet users. Horrigan (2007) combines in his approach to Internet user typologies a mix of the availability of technological assets, activities people engage in and their attitudes towards information and communication technologies (ICTs). His 10 types are grouped into three categories — elite tech users, middle of the road tech users and those with few tech assets, according mainly to the availability of technological devices in people's lives. Meyen and his colleagues (2010) analyze Internet users according to how they use the Internet to enhance capital. Meyen, *et al.* see the Internet both as a resource from which to get information as a way of enhancing cultural capital, and as a resource for social activity as a way of enhancing social capital. Meyen, *et al.* [1] distinguish seven different user types based on large number of qualitative interviews, identifying types according to their engagement in different capital–enhancing activities.

For the purposes of this paper, we have operationalized the Internet user classification based on users' primary online activities. The original approach to this Internet user typology (Pruulmann–Vengerfeldt, 2006) examined a long list of different activities online, applied factor analysis to to distil the basic underlying patterns and later applied cluster analysis to determine the key types of Internet users. Later the factor analysis was skipped and cluster analysis of the nationally representative data (in Estonia) and a long list of Internet related activities was repeated over several samples (Pruulmann–Vengerfeldt and Reinsalu, 2009). This confirmed that six basic user types, differing significantly based on their dominant Internet use principles, can be identified, as follows:

1. *Active versatile Internet users,* who are active in both information seeking, communication and entertainment related use;

2. *Practical work oriented Internet users*, who are mainly active information–oriented users;

3. *Entertainment oriented active Internet users,* whose main interests include entertainment and communication related uses, and seeking Internet solutions that cater to their interests;

4. *Practical information oriented small–scale Internet users* differ from the previous group in that while their activities focus mainly on information use, they are less frequent in their activities;

5. *Entertainment and communication oriented small–scale Internet users* also use the Internet less frequently than their active counterparts. Their use focuses on leisure–related activities and they are passive when it comes to information–related activities;

6. *Small–scale Internet users* use the technologies so infrequently that they do not have any significant types of activities that would describe them, and have very poorly developed online behavior.

This approach has yielded similar types to those of Meyen, *et al.* (2010). Versatile Internet Users correspond to Virtuosi, the Entertainment Oriented Active type to Addicts, the Practical Work Oriented type to Professionals. The less active types have in Meyen, *et al.'s* classification middle relevance to capital collection while the Practical Information Oriented small–scale users can be seen as Aficionados; Entertainment and Communication Oriented small–scale users correspond to Companions. In our previous studies, the group identified as the Small–scale Internet Users are in the Meyen, *et al.* study divided in two — the Cautious and the Affiliated. Despite the apparent similarities with the Meyen, *et al.* (2010) classification, we will use the longer, but more detailed, names that give slightly more nuanced descriptions of the activities.

In addition to these approaches there are studies that look at specific user groups, for example the classification of young internet users by Livingstone and Helsper (2007), Eyon and Malmberg (2011), or Holmes (2011). However, for the purpose of this study we needed a fairly simple classification system that allows users for self–classification with a minimum number of questions. This way we could support a relatively diverse range of Internet users. Therefore our study follows the ideas of the Kalmus, *et al.* (2009) and Meyen, *et al.* (2010) approach distinguishing between information–oriented and social–use oriented users, while at the same time identifying active and passive Internet users in both of these scales.

*Simple search, complex search, and exploratory search*

Simple or look–up searches are among the most basic types of search tasks. Usually they are present in the context of answering questions and fact finding (White and Roth, 2009). They are used to answer who, when and where questions. Users often show a query–answer behavior when carrying out simple tasks. A corresponding query can usually be formulated from a simple information need. When the user enters his or her query into a search system, the search results presented are, ideally, relevant to the query. In this paper, simple searches and look–up or fact–based searches are seen as being identical.

According to Campbell (1988), researchers disagree on what makes a task complex. On the one hand task complexity can be defined objectively, independent of the person carrying out the task. On the other hand it can also be defined by looking at the cognitive demands that are required from the performer of the task, which is totally dependent on the individual. Campbell (1988) proposes a classification for task complexity that is derived from a combination of the following factors influencing complexity: the presence of multiple paths to a desired end–state, the presence of multiple desired end–states, the presence of conflicting interdependence and presence of uncertainty or probabilistic linkages.

Singer, *et al.* (2012a) have presented a model that deconstructs the time–consuming complex search process into the steps aggregation, discovery and synthesis. According to Singer, *et al.* (2012a), a complex search is defined as a multi–step and time consuming process. A complex search is not answerable with one query, needing instead synthesized information from more than one retrieved Web page or document (Singer, *et al.*, 2012a). A complex search task is one that leads to a complex search activity. A (complex) search task is therefore described in relation to the (complex) search itself, which is an interactive process.

Exploratory search tasks are defined as open–ended information needs that are abstract and poorly defined and of a multifaceted character (Marchionini, 2006; White and Roth, 2009). Such exploratory search tasks fulfil needs such as learning, investigating or decision–making and require a high amount of interaction. They are usually ambiguous and uncertain and require a great deal of discovery.

A number of search tasks require a lot of effort but do not show all the standard attributes of exploratory search tasks, for example learning, planning and decision–making (White and Roth, 2009): They are therefore complex but not necessarily exploratory as defined before. Hence exploratory search tasks can be seen as a subset of complex search tasks.

**Research questions**

This study provides a good opportunity to scrutinize one dominant aspect of Internet use — search. We investigate the correlation between people's preferences in online behavior (applying the Internet user type concept) and their search performances. The assumption is that the success rate, speed and competence with which different user types search on the Internet differ. With the research that we present in this paper we have tried to answer the following research questions:

> RQ1: Is there a difference in search characteristics between certain user types when performing simple search tasks?
>
> RQ2: Is there a difference in search characteristics between certain user types when performing complex search tasks?
>
> RQ3: Is there a user–type specific difference in performance between simple and complex search tasks?

In general, Internet user types demonstrate different skill levels — the more active and more complex the user type's activities are, the higher their own rating of their skills (Pruulmann–Vengerfeldt and Reinsalu, 2009). This research enables us to show whether universal skill evaluation also holds true when conducting Internet searches. The performance questions also enable us to look deeper into specific characteristics of the search behavior and, based on our relatively moderate sample, compare these differences.

**Methods**

The results presented in this paper are based on a body of data gathered in the course of a larger experiment performed in August 2011. Two additional articles using distinct parts of the data and describing different aspects have been submitted for review so far (Singer, *et al.*, 2012b, Singer, *et al.*, 2012c). The following description of the research design is based on Singer, *et al.* (2012b). A more detailed outline about the experimental set–up can be obtained from there.

Rieger (2009) stresses the importance of the triangulation of data in order to understand search behavior. In our study, we also combine several methods. The experiment was conducted in a laboratory environment combining a self–evaluation survey, demographics and Internet user background data obtained from the survey, and search logs and results analysis. The user sample consisted of 60 volunteers, representing a cross–section of society in terms of age and gender (see Table 1 for further information on the participants). While we are well aware that such a sample is far from being representative, it is relatively large for a lab–based user study and gives a good in–depth insight into all user groups, with the exception of the small–scale Internet users.

The experiment was conducted in Germany, and the language of the search tasks was German. Participants were recruited in various ways (*e.g.*, through advertising). We aimed for an even distribution of participants concerning gender and age groups, although, this proved not completely possible. Even so, we think that the sample is good enough for our exploratory and experimental purposes. The sample consists of 32 women and 28 men, aged between 18 and 59 (see Table 1).

**Table 1: User sample.**

| Age | Female | Male | Total |
|---|---|---|---|
| 18–34 | 14 | 11 | 25 |
| 35–54 | 15 | 16 | 31 |
| 55–59 | 3 | 1 | 4 |
| Total | 32 | 28 | 60 |

The search example consisted of 12 tasks that could be classified as complex search tasks of varying complexity, as defined in the related work section. The condition of each task was that the answer had to be available somewhere on public Web sites in German as of August 2011. We set the sequence of tasks up in a way that users could alternately solve simple and complex tasks. The aim was to keep the participants interested, and to not discourage them through a sequence of complex search tasks which they might be unable to solve. The whole experiment consisted of six simple tasks indicated by (S) and six complex tasks indicated by (C), as described in Table 2.

| Table 2: Characteristics of simple and complex search tasks. | |
|---|---|
| **Characteristics of simple tasks (S)** | **Characteristics of complex tasks (C)** |
| closed tasks, information accessible with a single query, no browsing of Web sites required | open task, browsing required, multiple querying necessary, aggregating, discovery and synthesizing of steps necessary in the search process (see Kules and Capra, 2009; Singer, *et al.*, 2012a) |

Table 3 shows the differences in performance between 336 simple tasks (56 users times six simple tasks) and 336 complex tasks (56 users times six complex tasks) carried out during the experiment. While 73 percent of the simple tasks were correctly carried out, in only 45 percent of the complex tasks the users found the right result. Twenty–seven percent of the simple tasks were either partly correctly or wrongly carried out, or no result was submitted. In case of complex tasks this number was with 55 percent almost twice as high.

| Table 3: Performance differences between 336 simple and 336 complex search tasks in the experiment. | |
|---|---|
| **Simple tasks (S)** | **Complex tasks (C)** |
| 246 (73%) tasks correctly performed<br>6 (2%) tasks partly correctly performed<br>38 (11%) tasks wrongly performed<br>46 (14%) submitted without findings | 151 (45%) tasks correctly performed<br>62 (19%) tasks partly correctly performed<br>52 (15%) tasks wrongly performed<br>71 (21%) tasks submitted without findings |

Two exemplary task descriptions are:
(S) When and by whom was Penicillin invented?
(C) What are the five most important points to consider if you want to plan a budget wedding?

Before starting with the first search task, each study participant completed a short demographic survey. They had to indicate on a three–point scale how often they accessed the Internet per week (1–3 days, 4–6 days, every day) and on another three–point scale how long they were online per day (up to 2 hours, 2–5 hours, more than 5 hours). To analyze each respondent's Internet user type they were asked to indicate how regularly they engaged in certain Internet activities on a scale of 1 (never) to 5 (very often) as follows. Using summa indices we compared relative activity levels on the scale of information–related activities (information seeking for work purposes and information seeking for private purposes) and by entertainment–communication related activities (communicating with friends or family, work related communication, writing or commenting on forum entries, taking part in social networks, distributing their own content like videos, images, blog entries). As the scales were of different lengths, having two questions on information related activities and six questions on communication and entertainment related activities, we used a relative evaluation comparing each result to the possible maximum. If both scales were high, a person was considered an *active versatile Internet user type* (type 1). If the person was an active information user, but less active as entertainment–communication users, then the user type assigned was *practical work oriented Internet user type* (type 2). If the entertainment and communication related activities were more dominant and information related activities less, then the user type assigned was entertainment–oriented active Internet user (type 3). These types were regular everyday Internet users. If the overall user activities were less frequent, then the *practical information oriented small–scale user type* (type 4) or *entertainment and communication oriented small–scale Internet user type* (type 5) was assigned, depending on which activities were more dominant. There were no *small–scale Internet users*, as the laboratory style of the work meant that the few small–scale users recruited were unable to use the computer at a level sufficient to participate in the experiment. Table 4 shows a division of the respondent population according to Internet user type. For the sake of simplicity we will work with the Internet user type numbers instead of names for the rest of the paper. To get an indication of how well this sample represents the whole population, we can compare this sample to larger samples, such as a population–representative survey (*e.g.*, Pruulmann–Vengerfeldt and Reinsalu, 2009), which is of course country specific but still more expressive. The less active Internet user group is under-represented, especially as small–scale users did not participate. However, at the same time, each group is represented in order to give the possibility for preliminary statistical comparison (Table 4). The total number of the respondents comes to 56 (instead of the 60 study participants who took part in the experiment) as four respondents had to be excluded from the sample due to corrupt data.

| Table 4: Distribution of Internet user types in the study sample. | | | |
|---|---|---|---|
| Internet user type number | Internet user type name | Percentage of respondents in this type; number of users in this type | Percentage of representatives of this type in the nationally representative sample of the Estonian population (as a reference point) |
| 1 | Active versatile Internet users | 18%; n=10 | 10% |
| 2 | Practical work oriented Internet users | 25%; n=14 | 16% |
| 3 | Entertainment oriented active Internet users | 29%; n=16 | 13% |
| 4 | Practical information oriented small–scale Internet users | 14%; n=8 | 15% |
| 5 | Entertainment and communication oriented small–scale Internet users | 14%; n=8 | 10% |

We used the following set of standard measures in our analysis:

- Ranking: We created a ranking of users according to their overall performance in the experiment. We ranked the users first by the number of correct answers, and then, for users with the same number of correct answers, by the number of correct elements within answers;
- SERP time: The time (in seconds) users spent on search engine results pages (SERPs);
- Read time: The time (in seconds) users spent reading and checking Web pages (other than SERP);
- Task time: The time (in seconds) it took users to finish a task (not taking into consideration the outcome of the task);
- Number of opened tabs: Number of browser tabs users opened per search task;

- Number of queries: Number of queries users entered into search systems per task.

## Results

In this section we present the results of our experiment. We first state our findings regarding the correlation between user type and simple search tasks. We then show how the Internet user types perform when carrying out complex search tasks. And finally we work out the differences in participant behavior when confronted with simple and complex search tasks.

> RQ1: Is there a difference in search characteristics between certain user types when performing simple search tasks?

To answer this question, we have ranked the users according to their performance when carrying out simple search task only. The ranking was compiled by counting the number of correctly performed tasks and ranking users from high to low, then counting the number of partly correctly performed tasks and using this number as a second ranking factor. The users who scored better received a smaller rank number. The ranges for the ranks for the simple and complex tasks were significantly different. While for the simple tasks all users could be classified in ranks 1–10, the range was 1–18 for complex tasks. This is due to the fact that many more people managed to get the simple tasks right than people who managed to get the complex tasks right (20 people were ranked number 1 for the simple tasks but only 6 for the complex tasks). To make the two rankings comparable we have scaled the ranking numbers to 1–10 for simple and complex tasks.

As outlined in Table 5, the average rank for Internet user type 1 was 2.7 versus an average ranking of 4.8 for Internet user type 5. It seems that the Internet user type is an indicator for performance when carrying out simple search tasks.

Table 5: Ranking and search measures for simple search tasks (mean value ± standard error of mean).

| User type | Rank for simple tasks | SERP time (sec) | Read time (sec) | Number of tabs opened | Total task time (sec) | Number of queries |
|---|---|---|---|---|---|---|
| 1 | 2.7±0.9 | 31±9 | 88±15 | 3.8±0.6 | 119±23 | 2.0±0.3 |
| 2 | 3.4±0.7 | 27±5 | 125±17 | 6.4±1 | 152±20 | 1.9±0.2 |
| 3 | 3.8±0.8 | 40±9 | 109±19 | 3.5±0.4 | 149±27 | 2.2±0.3 |
| 4 | 4.8±1.3 | 26±10 | 91±14 | 6.7±2.2 | 116±20 | 1.8±0.6 |
| 5 | 4.8±1.2 | 41±11 | 109±15 | 3.2±0.5 | 150±23 | 2.4±0.4 |

The other search indicators, such as time spent on SERPs, time spent reading content pages, number of tabs opened, total task time and number of queries, also differ between groups. There is no clear direction to the differences because user types are non–hierarchical; in addition, and contrary to possible expectation, there is a lack of any pattern showing the more experience a person has, or the more specific the type of use practice, the more favored certain types of search behavior become.

However, although the difference between the mean values (in Table 5) seems obvious, statistical analysis showed no significant mutual differences between neighbouring pairs (user type 1 and 2, 2 and 3, 3 and 4 or 4 and 5), nor could we see significant difference of mean values when we grouped user types 1, 2 and 3 in one group and user types 4 and 5 into a second group. This can be explained by the relatively small sample size in each paired case. In terms of "number of tabs opened" though we found that the group consisting of types 1, 3 and 5 opens significantly fewer tabs than the group consisting of type 2 and type 4 (3.5±0.3 vs. 6.5±1.0 with a p value <0.01). This might be due to the fact that types 1, 3 and 5 are more entertainment oriented users whereas types 2 and 4 are more practical work and information oriented users.

Table 6: Spearman's *rho* (ρ) correlation coefficients between Internet user type and selected search measures for simple search tasks.

| | Spearman's *rho* | |
|---|---|---|

|  | (ρ) | p value |
|---|---|---|
| Rank | 0.24 | 0.03 (s) |
| SERP time | 0.20 | 0.20 (n.s.) |
| Read time | 0.01 | 0.49 (n.s.) |
| Number of tabs opened | -0.13 | 0.84 (n.s.) |
| Total task time | 0.07 | 0.31 (n.s.) |
| Number of queries | -0.07 | 0.70 (n.s.) |

Finally we investigated the correlation between Internet user types and search measures, as outlined in Table 6. As the Internet user type is an ordinal variable we have used Spearman's *rho* correlation coefficients to express the correlation between the variables. In line with the previous discussion, the strongest correlation is between ranking number and Internet user type, and SERP time and Internet user type. Of the correlation coefficients stated in Table 6, only the ranking number is statistically significant (indicated by s) with a p value of 0.03. The correlation measures for the remaining parameters are weak and will not be discussed further.

> RQ2: Is there a difference in search characteristics between certain user types when performing complex tasks?

We have ranked the users according to their performance when carrying out complex search tasks only. The ranking was compiled the same way as explained in the previous section. As outlined in Table 7, the average rank for Internet user type 1 was 4.9 versus an average rank of 6.0 for Internet user type 5. It seems that Internet user type is an indicator for performance when carrying out complex search tasks. However, Internet user type 5 shows a better average rank than Internet user type 4.

| Table 7: Spearman's *rho* (ρ) correlation coefficients between Internet user type and selected search measures for simple search tasks. | | | | | | |
|---|---|---|---|---|---|---|
| User type | Rank | SERP time (sec) | Read time (sec) | Number of tabs opened | Total task time (sec) | Number of queries |
| 1 | 4.9±1 | 83±13 | 264±34 | 3.9±0.7 | 347±36 | 6.1±1.2 |
| 2 | 4.4±0.7 | 112±14 | 349±47 | 3.8±0.6 | 461±49 | 5.6±0.4 |
| 3 | 4.8±0.8 | 124±15 | 260±30 | 2.7±0.4 | 385±37 | 6.1±0.8 |
| 4 | 6.9±0.8 | 135±37 | 329±62 | 2.9±0.6 | 464±83 | 7.7±2.2 |
| 5 | 6.0±1.2 | 168±42 | 359±52 | 3.2±0.8 | 527±81 | 7.4±1.9 |

Table 7 also shows that some of the other criteria that we used to evaluate search behavior differ among our respondents. Although in this experiment we have placed 'getting the answer right' as the most important ranking criterion, the differences in speed, number of queries and skill in using multiple browser tabs to compare different results can also be interpreted as indicators of search skills. However, our study does not indicate that there is a single 'winning' formula to getting the best search results. Rather, our results indicate that being able to switch search strategies by adding keywords or opening additional browser tabs if required may be an important skill.

Next we checked the differences of the mean values presented in Table 7 for significance. We could not demonstrate significant mutual differences between neighbouring pairs (Internet user type 1 and 2, 2 and 3, 3 and 4 or 4 and 5). However, there is a significant difference of mean values when we group active user types (1, 2 and 3) in one group, and small–scale or more passive user types (4 and 5) in another (p <0.03). This means that the active and more experienced Internet users perform significantly better when conducting complex search tasks than the group consisting of small–scale Internet users.

We also grouped Internet user types focused on information (types 2 and 4) and compared them with the group of Internet users whose main activities relate to communication and entertainment (types 3 and 5), but did not see a statistically significant difference (p>0.5) in the average rankings of the two groups. This means that neither of the two groups performed better than the other. Compared to the preceding discussions, this shows that experience and competence obtained from regular and intensive use can be more decisive factors than what a person does online. This might need further investigation using a larger sample, although in this study the idea that activity type is significant was not confirmed.

**Table 8: Spearman's *rho* (ρ) correlation coefficients between Internet user type and selected search measures for complex search tasks.**

|  | Spearman's *rho* (ρ) | p value |
|---|---|---|
| Rank | 0.22 | 0.06 (n.s.) |
| SERP time | 0.25 | 0.03 (s) |
| Read time | 0.11 | 0.21 (n.s.) |
| Number of tabs opened | -0.19 | 0.92 (n.s.) |
| Total task time | 0.18 | 0.09 (n.s.) |
| Number of queries | 0.002 | 0.51 (n.s.) |

Finally we investigated the correlation between Internet user type and other search measures, as outlined in Table 8. As previously in the case of simple tasks, we again used Spearman's *rho* correlation coefficients to express the correlation between the variables. Again the rank correlated positively with Internet user type, as also observed in the case of simple tasks. In this case, there are also other measures that correlate, for example time spent on search engine results pages, read time and total search time. The number of browser tabs opened correlated negatively. No correlation could be found for the number of queries. Of the correlation coefficients stated in Table 8, only the SERP time is statistically significant (indicated by s) with a p value of 0.03. The correlation measures for the remaining parameters are weak and will not be discussed further.

> RQ3: Is there a user–type–specific difference in performance between simple and complex tasks?

In the case of complex tasks, the average rank is significantly lower (indicating worse performance) for all user types in comparison to simple tasks as outlined in Table 9. In the case of complex tasks the rank difference between Internet user type 1 and Internet user type 4 is 2.0, whereas in the case of simple tasks the difference is 2.1 (type 1 and 4).

**Table 9: Ranking for simple and complex tasks (scale 1–10, 1 being the highest rank).**

| User type | Average rank for simple tasks (1=highest) | Average rank for complex tasks (1=highest) |
|---|---|---|
| 1 | 2.7±0.9 | 4.9±1 |
| 2 | 3.4±0.7 | 4.4±0.7 |
| 3 | 3.8±0.8 | 4.8±0.8 |
| 4 | 4.8±1.3 | 6.9±0.8 |
| 5 | 4.8±0.7 | 6.0±1.2 |

We have run paired–sample t–tests comparing complex vs. simple rankings for each user type. The resulting p values (>0.5) indicate that the difference between the mean values is not significant. As our user sample was comparably small (N=60) and the users had varying levels of search experience (ranging from inexperienced housewives to experienced students of information science), we assume that a larger sample in combination with a more homogeneous average search experience would lead to smaller standard errors and clearer results.

### Discussion

As the amount of information available on the Internet is steadily rising, the skill of finding necessary information online becomes more and more relevant (Weinreich, *et al.*, 2008; Aula, *et al.*, 2005; Hölscher and Strube, 2000). Our comparative study of Internet user types indicates that search engines are doing a fairly good job in supporting simple query–based searches, as confirmed by the fact that 73 percent of the simple tasks but only 45 percent of the complex tasks in our experiment were correctly carried out. This is also confirmed by the research by Rieger (2009).

When going into more detail about the correlation between user type and search performance, our findings show that the user type generally correlates with the search performance. In the case of simple tasks people who are more active on the Internet and regularly perform various kinds of activities from information seeking over communicating to entertainment related activities (both in their leisure time as well as professionally) perform significantly better than their less active counterparts (confirmed by the significant correlation coefficient of 0.24 between rank number and Internet user type for simple tasks in Table 10).

In the case of complex tasks this difference in performance between more and less active user types does not increase as one would expect but even slightly decreases (confirmed by a weakly significant correlation coefficient between rank number and user type of 0.22 in Table 10).

| | **Table 10: Spearman's *rho* (ρ) correlation coefficients between Internet user type and selected search measures for simple and complex search tasks.** | | | |
|---|---|---|---|---|
| | **Simple tasks** | | **Complex tasks** | |
| | Spearman's *rho* | p value | Spearman's *rho* | p value |
| Rank | 0.24 | 0.03 (s) | 0.22 | 0.06 (n.s.) |
| SERP time | 0.20 | 0.20 (n.s.) | 0.25 | 0.03 (s) |
| Read time | 0.01 | 0.49 (n.s.) | 0.11 | 0.21 (n.s.) |
| Number of tabs opened | -0.13 | 0.84 (n.s.) | -0.19 | 0.92 (n.s.) |
| Total task time | 0.07 | 0.31 (n.s.) | 0.18 | 0.09 (n.s.) |
| Number of queries | -0.07 | 0.70 (n.s.) | 0.002 | 0.51 (n.s.) |

Previous user experience plays a slightly more important role for simple tasks than for complex tasks. We explain this with the fact that all users in the experiment (and hence all user types from more active to less active) have relatively more previous experience with carrying out simple tasks than with complex tasks. When comparing Tables 5 and 7 the ranking numbers also confirm that both groups equally struggle for complex tasks — confirmed by a rank range from 2.7 to 4.8 for simple tasks and 4.9 to 6.9 for complex task. The difference between the best ranking type and worst ranking type is comparable for simple tasks (difference of 2.1 between type 1 and type 5) as for complex tasks (difference of 2.0 between type 1 and type 4).

Overall, user types 4 and 5 spend more time on search engine results pages looking for the right answers, more search sessions are needed. The interpretation could be that searchers of types 4 and 5 seem to expect the search engine to provide the correct answer and are less likely to use a comparison of multiple sources in order to deduce the right answer. This idea seems to be supported by the fact that the Internet user type correlates negatively with the number of browser tabs opened. This might be an indicator that users who are less active online, are also less familiar with using tabs to browse and search more efficiently or to compare different results.

The Internet user type comparison shows differences only when looking at the more active (types 1 to 3) and passive user types (types 4 and 5), although the assumption that there would also be differences when comparing information–related users and communication–related users was not confirmed in this study. Further studies with more specific user groups could be performed with search tasks tailored to measure both, success in seeking information and entertainment and communication.

The results of our studies also have implications for formal and non–formal learning as the overall results show that people need support when learning how to use different search strategies, and especially when learning how to manage complex search tasks. Despite the fact that our study did not provide a single 'winning' formula for the most successful search strategies, there is enough information to expand the analysis to look for different search patterns and strategies, which could later be recommended for introduction to education programs. Special support is needed for those with less experience in Internet related activities as they tend to prerform poorer in the overall evaluation.

## Limitations and conclusion

Since the invention of the Internet and its rapid growth over the last decade, users carry out more and more of their daily routine tasks over the Web. An essential activity when being online is looking for and finding information. In this paper we presented the results of a study examining whether a person's Internet user type, a classification of what people do online (be it communicating with friends, blogging, spending time on social networking sites like Facebook or writing entries on forums), correlates with their ability to search online. The user sample of 56 ordinary Web users is rather special and differentiates it from a long list of user studies, which are carried out with quite narrow user groups (for example university students).

Our findings are that in the case of complex tasks active Internet users (types 1, 2 and 3 grouped together) show a significantly different searching behavior to more small–scale Internet users (types 4 and 5 grouped together).

As far as search engine vendors are concerned, we think that it would make sense to assign an Internet user type to each search engine user. Based on this user type, each user could be offered Internet–specific support (like recommendations to use different tabs to compare results or suggestions to modify search terms either to be more specific or to be broader). To a certain extent this is happening already as the degree of personalization from the search engine side is increasing (Feuz, *et al.*, 2011; Zimmer, 2008), although analyzing the user behavior would also mean recommending databases and for example encyclopaedia pages that are out of the 'normal' pathway, going seemingly against the idea of algorithms proposing the more frequented pages. This, however, also means semantic analysis of questions, and search engines would need to be able to distinguish between simple and complex searches.

As far as the limitations of our study are concerned, it is, of course, possible that users show different behavior in the laboratory environment than they would show in real life. While search engines seem to be the obvious solution in a laboratory experiment situation, entertainment and communication related users as well as the active versatile users might in other situations turn to their friends and peers in social networks.

Another limitation is the sample used in our study. It was our intention to use a selection of users of a wide age span, and also men and women alike, so that the results would in some sense represent society as a whole. This variety of users resulted in quite high variation of the measures we analyzed, and sometimes the differences between averages (*e.g.*, between certain user types) was not as clear as expected.

The time limit that users were faced with is also a limitation. Some users asked to be given more time after the three hours reserved for the experiment. Therefore, we assume that study participants would have (successfully) completed more tasks in the case of an open–ended experiment or a long–term study in a naturalistic environment (Kelly, 2009; Singer, *et al.*, 2011). On the other hand it is well known that users put more effort into completing search tasks in a lab setting (Rubin and Chisnell, 2008). This limitation is not unique to our study.

In future experiments, we are planning to use more nuanced user samples that allow us to compare for example two specific user types with each other and work out their differences. These comparisons should provide the possibility to zoom in on activities. The analysis in this paper takes a statistical birds–eye view of the search process, while the data and data collection method enables us to investigate actual search patterns and search strategies. The browser based logging software would also enable us to follow the user in her or his natural search context and investigate the searches occurring naturally over the course of a day for a given user type. That data–rich natural experiment would give non–commercial tracking information and enable us to see search patterns carried across different Web sites and different periods of time. Supporting this with other methods (*e.g.*, diary) would also enable us to look at cross–media search in an attempt to understand the searching for information in the context of other sources. The search logger software would simplify this kind of cross–media approach as the search can be connected to other activities in the Internet browser.

## About the authors

**Georg Singer** is a researcher and lecturer at the University of Tartu in Estonia. In his Ph.D. he tries to improve the support for complex Web searching in current search engines. Prior to this, he has worked in various international management positions. He is currently also active as a Web entrepreneur.
Direct comments to georg [dot] singer [at] ut [dot] ee

**Pille Pruulmann–Vengerfeldt** is an associate professor at the University of Tartu, Institute of Journalism and Communication, and a researcher at the Estoniian National Museum in Estonia. Her interests are Internet user typologies, user–friendly online spaces as possible venues for participation and participatory applications for organisations. She is leader of the research project 'Developing Museum Communication in the 21st Century Information Environment' and partner in the FP7 project 'Usable and Efficient Secure Multi–party Computation'. She is also a participant in the 'EU Kids Online' and 'Estonia as an Emerging Information and Consumer Society: Social Sustainability and Quality of Life' projects. She has recently been published in the *Journal of Baltic Studies*, *Journal of Computer–Mediated Communication* and the *Journal of Children and Media*.

**Ulrich Norbisrath** is a senior research fellow at the University of Tartu, Estonia. He has been working in Estonia since the end of 2006. He teaches courses on agile and visual software development and systems administration. He is involved in research projects from the areas of ubiquitous computing and information management. His research groups are part of the Distributed Systems group in Tartu. He invented the concept of Friend–to–Friend (F2F) Computing (http://f2f.ulno.net) and is one of the founders of the Search–Logger project (http://www.search-logger.com). He is currently working on creating a biometric technologies broker framework, and researching new ways to support complex search. In his Ph.D. (2006) at the RWTH Aachen University (Aachen, Germany) he researched a way to enable the easy deployment of smart home services in random physical environments.

**Dirk Lewandowski** is a professor of information research and information retrieval at the Hamburg University of Applied Sciences. Prior to that, he worked as an independent consultant and as a part–time lecturer at the Heinrich Heine University, Düsseldorf. His research interests are Web Information Retrieval and users' interaction with Web search engines.

## Acknowledgements

This research was partly supported by the Estonian Information Technology Foundation (EITSA) and the Tiger University program, as well as by Archimedes and Target financing project SF0180017s07.

## Note

1. Meyen, *et al.*, 2010, p. 876.

Ryen W. White, Gary Marchionini, and Gheorghe Muresan, 2008. "Editorial: Evaluating exploratory search systems," *Information Processing and Management*, volume 44, number 2, pp. 433–436. http://dx.doi.org/10.1016/j.ipm.2007.09.011

Michael Zimmer, 2008. "The externalities of search 2.0: The emerging privacy threats when the drive for the perfect search engine meets Web 2.0," *First Monday*, volume 13, number 3, at http://firstmonday.org/article/view/2136/1944, accessed 15 March 2012.

---

**Editorial history**

Received 16 February 2012; revised 31 March 2012; revised 13 May 2012; revised 21 May 2012; accepted 14 May 2012.

---